\documentclass[prb,twocolumn,showpacs,superscriptaddress,preprintnumbers,amssymb]{revtex4}
\usepackage{graphicx}
\usepackage{dcolumn}
\usepackage{bm}
\usepackage{wasysym}

\newcommand{\beq}{\begin{equation}}
\newcommand{\eeq}{\end{equation}}
\newcommand{\beqn}{\begin{eqnarray}}
\newcommand{\eeqn}{\end{eqnarray}}

\begin{document}

\title{Phase Diagram of the Kane-Mele-Hubbard model}

\author{Christian Griset}


\author{Cenke Xu}

\affiliation{Department of Physics, University of California,
Santa Barbara, CA 93106}

\date{\today}

\begin{abstract}

Motivated by recent numerical results, we study the phase diagram
of the Kane-Mele-Hubbard (KHM) model, especially the nature of its
quantum critical points. The phase diagram of the
Kane-Mele-Hubbard model can be understood by breaking the
$\mathrm{SO(4)}$ symmetry of our previous work down to
$\mathrm{U(1)_{spin} \times U(1)_{charge} \times PH}$ symmetry.
The vortices of the inplane N\'{e}el phase carry charge, and the
proliferation of the {\it charged} magnetic vortex drives the
transition between the inplane N\'{e}el phase and the QSH
insulator phase; this transition belongs to the 3d XY universality
class. The transition between the liquid phase and the inplane
N\'{e}el phase is an anisotropic O(4) transition, which eventually
becomes first order due to quantum fluctuation. The liquid-QSH
transition is predicted to be first order based on a $1/N$
calculation.

\end{abstract}
\pacs{} \maketitle

\section{Introduction}

Thanks to the discovery of Graphene
\cite{graphene1,graphene2,graphene3}, a great deal of attention
has been devoted to systems with Dirac fermions at low energy. It
was demonstrated that many topological states of fermions are
related to Dirac fermions, such as the quantum Hall state
\cite{haldane1988}, quantum spin Hall state
\cite{kane2005a,kane2005b}, and 3d Topological insulator
\cite{kane2007,fukane2007}, etc. Since last year, motivated by the
quantum Monte Carlo (QMC) simulation on the Hubbard model on the
honeycomb lattice \cite{meng}, strongly interacting Dirac fermions
have stimulated a lot of interests. Quite unexpectedly, a fully
gapped liquid phase was discovered in the phase diagram of the
honeycomb lattice Hubbard model at intermediate Hubbard $U$
\cite{meng}, and by increasing $U$ this liquid phase is driven
into a N\'{e}el phase after a continuous quantum phase transition.
This liquid phase has stimulated many theoretical and numerical
studies on possible spin liquid phases on the honeycomb lattice
\cite{ran2010,ran2010b,wang2010,clark2010,thomalehoney,xusachdev2010,xu2010,taoli2010}.
So far almost all the theoretical proposals about this liquid
phase involve nontrivial topological orders
\cite{ran2010,ran2010b,wang2010,xusachdev2010,xu2010}.

The Hubbard model on the honeycomb lattice has the full
$\mathrm{SO(4)\sim [SU(2)_{spin} \times SU(2)_{charge}]}/Z_2$
symmetry \cite{yangzhang,zhang1991}, thus a true liquid phase of
the Hubbard model should preserve all these symmetries. In Ref.~
\cite{xusachdev2010,xu2010}, a full SO(4) invariant theory of the
Hubbard model was developed, and it was proposed that the liquid
phase observed in Ref.~\cite[]{meng} is a topological spin-charge
liquid phase with mutual semion statistics between gapped spin-1/2
and charge$-e$ excitations. The global phase diagram of this
theory is depicted in Fig.~\ref{honeypd}. We will review this
theory in the next section.

In the current work, we will consider perturbations on the Hubbard
model that break the SO(4) symmetry down to its subgroups. In
particular we will focus on the Kane-Mele-Hubbard model that was
recently studied numerically
\cite[]{qshhubbard1,qshhubbard2,qshhubbard3}: \beqn H &=&
\sum_{<i,j>, \alpha} -t c^\dagger_{i,\alpha}c_{j,\alpha} +
\sum_{\ll i,j \gg,\alpha,\beta} \lambda \ i\nu_{i,j}
c^\dagger_{i,\alpha}\sigma^z_{\alpha\beta} c_{j,\beta} \cr\cr &+&
Un_{i,\uparrow}n_{i,\downarrow}.  \label{KMH} \eeqn The second
term of this Hamiltonian is the spin-orbit coupling introduced in
the original Kane-Mele model for the quantum spin Hall effect
(QSH) \cite{kane2005a,kane2005b}. The goal of the current work is
to understand the change of the phase diagram and quantum critical
points due to the existence of the QSH spin-orbit coupling,
compared with the SO(4) invariant case.

\section{$\mathrm{SO(4)}$ invariant theory}

In Ref.~\cite[]{xu2010}, the author used the SO(4) symmetry to
classify the order parameters on the honeycomb lattice
\cite{xu2010}. In particular, the quantum spin Hall (QSH) and
triplet-superconductor (TSC) order parameters belong to a
$(\mathbf{3}, \mathbf{3})$ matrix representation of the SO(4)
group: \beqn && Q_{ab} = \left(
\begin{array}{cccc}
\mathrm{Im}(\mathrm{TSC})_x \ , & \mathrm{Im}(\mathrm{TSC})_y \ , & \mathrm{Im}(\mathrm{TSC})_z \\ \\
\mathrm{Re}(\mathrm{TSC})_x \ , & \mathrm{Re}(\mathrm{TSC})_y \ , & \mathrm{Re}(\mathrm{TSC})_z \\
\\ \mathrm{QSH}_x \ , & \mathrm{QSH}_y \ , & \mathrm{QSH}_z
\end{array}
\right) \label{Q}\eeqn  Since all these order parameters are
topological, their topological defects carry nontrivial quantum
numbers. For instance, a Skyrmion of the spin vector
$(\mathrm{QSH_x, QSH_y, QSH_z })$ carries charge$-2e$
\cite[]{senthil2007}, while a Skyrmion of the charge vector
$(\mathrm{Im(TSC)_z, Re(TSC)_z, QSH_z })$ carries spin$-1$ $i.e.$
spin and charge are dual to each other, and view each other as
topological defects \cite{xu2010}. The liquid phase proposed in
Ref.~\cite[]{xusachdev2010,xu2010} was obtained by proliferating
both the spin and charge Skyrmions from the condensate of
$Q_{ab}$.

In the SO(4) invariant theory, the low energy physics of
topological defects of the matrix order parameter $Q_{ab}$ is
described by the following theory: \cite{xusachdev2010,xu2010}:
\beqn \mathcal{L}_{cs} &=& \frac{2i}{2\pi}\epsilon_{\mu\nu\rho}
A^z_{c,\mu}\partial_\nu A^z_{s,\rho} \cr\cr &+& |(\partial_\mu - i
A^z_{s, \mu})z^{s}_\alpha|^2 + r_s|z^{s}_\alpha|^2 \cr\cr &+&
|(\partial_\mu - i A^z_{c, \mu})z^{c}_\alpha|^2 +
r_c|z^{c}_\alpha|^2 + \cdots \label{lcs} \eeqn The CP(1) fields
$z^s_\alpha$ and $z^c_{\alpha}$ are $\mathrm{SU(2)_{spin}}$ and
$\mathrm{SU(2)_{charge}}$ fundamental doublets. In terms of
$z^s_\alpha$ and $z^c_\alpha$, the order parameter $Q_{ab}$ is
represented as \beqn Q_{ab} \sim (z^{s \dagger} \sigma^a z^s)
(z^{c \dagger} \sigma^b z^c). \eeqn The mutual Chern-Simons field
in Eq.~\ref{lcs} identifies $z^s_{\alpha}$ ($z^c_\alpha$) as the
vortex (meron) of charge (spin) sectors of the order parameter
$Q_{ab}$.

\begin{figure}
\begin{center}
\includegraphics[width=2.6 in]{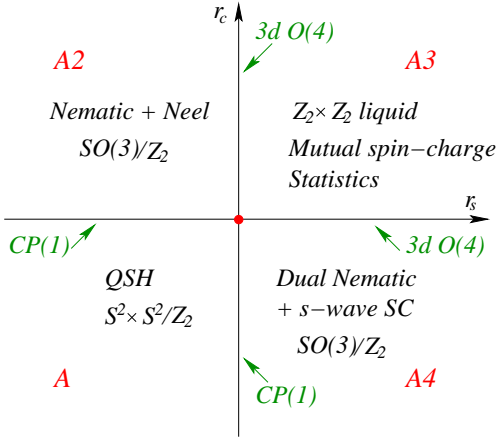}
\caption{The global phase diagram for the SO(4) invariant theory
Eq.~\ref{lcs} proposed in Ref.~\cite{xu2010}.} \label{honeypd}
\end{center}
\end{figure}

By tuning the parameters $r_s$ and $r_c$, a global phase diagram
Fig.~\ref{honeypd} is obtained. There are in total four phases:

{\it (1).} Phase $A3$ corresponds to the case with $z^s$ and $z^c$
both gapped, and the system is in a topological liquid phase with
mutual anyon statistics between spin$-1/2$ and charge$-e$
excitations. This statistics is guaranteed by the mutual CS fields
in Eq.~\ref{lcs}. In Ref.~\cite{xu2010} it was proposed that this
is the liquid phase observed by QMC \cite{meng}.

{\it (2).} In Phase $A2$, $z^s$ is condensed while $z^c$ is
gapped. The system is in a magnetic ordered phase with both
N\'{e}el and transverse nematic order. The ground state manifold
(GSM) of this phase is $\mathrm{SO(3)}/Z_2$.

{\it (3).} In phase $A$, both $z^s$ and $z^c$ are condensed, the
system is described by the condensate of order parameter $Q_{ab}$,
with GSM $(S^2\times S^2)/Z_2$. One example state of this phase is
the $\mathrm{QSH}_z$ state, which couples to the fermions in the
same way as the spin-orbit coupling introduced in the Kane-Mele
model \cite{kane2005a,kane2005b}.

{\it (4).} Phase $A4$ is the charge-dual of the phase $A2$, with
the same GSM $\mathrm{SO(3)}/Z_2$.

All the phase transitions in phase diagram Fig.~\ref{honeypd} are
continuous. For example, the transition between $A3$ and $A2$
belongs to the 3d O(4) universality class, while the transition
between $A$ and $A2$ is a CP(1) transition, which is equivalent to
the deconfined quantum critical point
\cite[]{deconfinecriticality1,deconfinecriticality2}. The
multicritical point in Fig.~\ref{honeypd} was studied using a
$1/N$ expansion in Ref.~\cite{xusachdevtriangle}, and when $N$ is
sufficiently large this multicritical point is a conformal field
theory.

The same field theory Eq.~\ref{lcs} was used to describe various
phases observed experimentally on the triangular lattice
frustrated magnets \cite{xusachdevtriangle}, such as
$\kappa\mathrm{-(ET)_2Cu_2(CN)_3}$,
$\mathrm{EtMe_3Sb[Pd(dmit)_2]_2}$,
$\mathrm{EtMe_3P[Pd(dmit)_2]_2}$, etc. Also, a similar theory
without $\mathrm{SU(2)_{charge}}$ was applied to the cuprates
\cite{yepeng2010,kouqiweng}.

\section{The Kane-Mele-Hubbard model}

\subsection{General formalism}

In the Kane-Mele-Hubbard (KMH) model Eq.~\ref{KMH}, the symmetry
of the Hubbard model is broken down to \beqn \mathrm{U(1)_{spin}
\times U(1)_{charge}} \times \mathrm{PH}, \eeqn where
$\mathrm{U(1)_{spin}}$ is the spin rotation around $z$ axis, while
$\mathrm{U(1)_{charge}}$ corresponds to the ordinary charge U(1)
rotation. The extra particle-hole symmetry (PH) is \beqn
\mathrm{PH} : c_{i,\alpha} \rightarrow c^\dagger_{i,\alpha}(-1)^i.
\label{PH} \eeqn This PH symmetry is in fact a product of spin and
charge $\pi-$rotations around the $y$ axis: $\mathrm{PH} =
\pi^{y,c} \cdot \pi^{y,s}$. One can verify that $\pi^{y,c}$ and
$\pi^{y,s}$ individually changes the KMH model, while their
product keeps the model invariant. The definition of PH is not
unique. For instance, one can also define PH as $\mathrm{PH} =
\pi^{x,c}\cdot \pi^{x,s}$, then fermion operator transforms as
$c_{i} \rightarrow \sigma^z c^\dagger_i (-1)^i$.

The QSH spin-orbit coupling corresponds to $Q_{33}$ of the matrix
order parameter $Q_{ab}$ in Eq.~\ref{Q}. Thus a nonzero $\langle
Q_{33} \rangle$ in the Hamiltonian will modify the field theory
Eq.~\ref{lcs} as follows: \beqn \mathcal{L}_{cs} &=&
\frac{2i}{2\pi}\epsilon_{\mu\nu\rho} A^z_{c,\mu}\partial_\nu
A^z_{s,\rho} \cr\cr &+& |(\partial_\mu - i A^z_{s,
\mu})z^{s}_\alpha|^2 + r_s|z^{s}_\alpha|^2 \cr\cr &+&
|(\partial_\mu - i A^z_{c, \mu})z^{c}_\alpha|^2 +
r_c|z^{c}_\alpha|^2 \cr\cr &+& u \langle Q_{33} \rangle (z^{s
\dagger} \sigma^z z^s) (z^{c \dagger} \sigma^z z^c) + \cdots
\label{lcs1} \eeqn Notice that terms like $z^{c \dagger} \sigma^z
z^c$, $z^{s \dagger} \sigma^z z^s$ etc. are all forbidden by the
PH symmetry. The symmetry-breaking introduced by the QSH
spin-orbit coupling will not change the nature of the liquid phase
($A3$ of Fig.~\ref{honeypd}). However, the other phases with spin
and charge orders will be modified. The modified global phase
diagram is depicted in Fig.~\ref{KMHphase}$a$.

\subsection{N\'{e}el phase and charged vortex}

\begin{figure}
\begin{center}
\includegraphics[width=3.5 in]{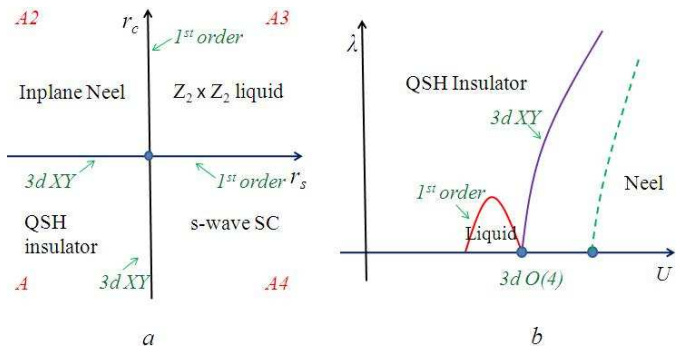}
\caption{$(a)$. The global phase diagram of Eq.~\ref{lcs1}, for
models that break SO(4) to $\mathrm{U(1)_{spin} \times
U(1)_{charge}} \times \mathrm{PH}$ symmetry. $(b)$. The phase
diagram of the actual KMH model. The theory in
Ref.~\cite{ran2010b} would predict an extra transition line inside
the N\'{e}el phase (dashed line), which corresponds to the
order-disorder transition of the CAF order parameter. This dashed
line is absent in our theory.} \label{KMHphase}
\end{center}
\end{figure}

Let us start with phase $A2$. With the background QSH order
parameter, the phase $A2$ of Fig.~\ref{honeypd} is reduced to a
pure inplane N\'{e}el phase with GSM $S^1$ in
Fig.~\ref{KMHphase}$a$.

One very special property of this N\'{e}el order is that, the
vortex of the N\'{e}el order carries unit electric charge due to
the ``dual" QSH effect, since the vortex of the inplane N\'{e}el
order carries a magnetic $\pi-$flux. This ``dual" QSH effect and
charged spin-flux was discussed in Ref.~\cite{ranvison,qivison}.
The key question we want to address here is, does the charge
carried by the vortex affect the quantum phase transitions around
the N\'{e}el phase?

This problem can be understood by classifying the charged vortices
of the N\'{e}el order based on the symmetry of the system. Every
charged vortex carries two quantum numbers, charge and vorticity,
denoted as $(e, v)$. There are in total four flavors of
charged-vortices: \beqn && z^c_1 = (e, v), \ \ \ (z^{c}_1)^\ast =
(-e, -v), \cr\cr && z^c_2 = (- e, v), \ \ \ (z^{c}_2)^\ast = (e,
-v).  \eeqn $z^c_1$ and $z^c_2$ are precisely the
$\mathrm{SU(2)_{charge}}$ doublet introduced in Eq.~\ref{lcs}. If
there is a full SO(4) symmetry, all four flavors of vortices are
degenerate. Within the current KMH model, the symmetry guarantees
that $z^c_\alpha$ and $(z^c_\alpha)^\ast$ are degenerate, but
$z^c_1$ and $z^c_2$ are {\it not} necessarily degenerate. This is
due to the fact that under the PH transformation in Eq.~\ref{PH},
both $e$ and $v$ are reversed. (Notice that PH defined in
Eq.~\ref{PH} transforms $S^x \rightarrow -S^x$, $S^y \rightarrow
S^y$.)

The classification of vortices can also be understood from the
general theory Eq.~\ref{lcs1}. For instance, in phase $A2$
(condensate of $z^s_\alpha$), due to the existence of the last
term of Eq.~\ref{lcs}, the condensate of $z^s_\alpha$ splits the
degeneracy between $z^c_1$ and $z^c_2$.

This inplane N\'{e}el order is always accompanied with a
background QSH spin-orbit coupling. Since the QSH spin-orbit
coupling breaks the reflection symmetry $x \rightarrow -x$ of the
honeycomb lattice, one might expect that the following spin order
parameter automatically acquires a nonzero expectation value:
\beqn H^\prime \sim \sum_{\ll i, j \gg} \nu_{ij} \hat{z}\cdot (
\vec{S}_i \times \vec{S}_j). \eeqn However, $H^\prime$ is {\bf
odd} under PH. Thus unless the system further breaks the PH
symmetry, $H^\prime$ should {\it not} have any nonzero expectation
value.

\subsection{N\'{e}el-QSH transition}

Usually the order-disorder transition of inplane XY order is
driven by proliferating the vortices of the order parameter. In
the previous section we have classified the vortices in the
inplane N\'{e}el phase. In the N\'{e}el phase, since vortices
$z^c_1$ and $z^c_2$ are not degenerate, only one component of the
vortex doublet $z^c_\alpha$ condenses at the transition. Let us
take this vortex to be $z^c_1$, then the field theory of the
transition is \beqn \mathcal{L} = |(\partial_\mu -
iA^z_{c,\mu})z^c_1|^2 + r_c|z^c_1|^2 + \cdots \eeqn This is a 2+1d
Higgs transition, which belongs to the 3d XY universality class.
The gauge field $A^z_{c,\mu}$ is precisely the dual of the
Goldstone mode of
the inplane N\'{e}el phase. 
The condensate of $z_1$ has no Goldstone mode due to the Higgs
mechanism, thus the condensate has no superconductor order even
though $z^c_1$ carries charge. The condensate of $z^c_1$ is
precisely the QSH insulator.

If we start with the QSH insulator phase, this QSH-N\'{e}el phase
transition can be viewed as condensation of magnetic exciton $b_i
\sim c^\dagger_{\uparrow,i}c_{\downarrow,i}$ \cite{dhlee2011}.
Since $(b_i)^2 = 0$, the magnetic exciton is a hard-core boson.
Under PH transformation, $b_i$ transforms as $b_i \rightarrow
-b_i^\ast$. This symmetry rules out the linear time-derivative
term in the Lagrangian of $b$, thus this transition is an ordinary
3d XY transition, which is consistent with the analysis in the
previous paragraph.

As a comparison to the KMH model, let us discuss a slightly
different kind of symmetry breaking of the Hubbard model. In this
case, the SO(4) symmetry of the Hubbard model is broken down to
$\mathrm{U(1)_{spin} \times U(1)_{charge}} \times \pi^{y,c} \times
\pi^{y,s}$, $i.e.$ both $\pi^{y,c}$ and $\pi^{y,s}$ are symmetries
of the system individually (in the KMH case only their product is
the symmetry). According to the symmetry $\mathrm{U(1)_{spin}
\times U(1)_{charge}} \times \pi^{y,c} \times \pi^{y,s}$, in the
inplane N\'{e}el phase (phase $A2$) all four flavors of charged
vortices (merons) with quantum numbers $(\pm e, \pm v)$ are
degenerate. Thus the low energy field theory describing these
charged vortices is the CP(1) model with easy-plane anisotropy:
\beqn \mathcal{L} &=& \sum_{\alpha = 1}^2 |(\partial_\mu -
iA^z_{c,\mu})z^c_\alpha|^2 + r_c|z^c_\alpha|^2 + g (\sum_{\alpha =
1}^2 |z^c_\alpha|^2)^2 \cr\cr &+& u |z^c_1|^2|z^c_2|^2 + \cdots
\eeqn We keep $u < 0$, thus the $\mathrm{SU(2)_{charge}}$ is
broken down to the easy-plane direction, $i.e.$ $z^c_1$ and
$z^c_2$ both condense at the transition. When $z^c_\alpha$ both
condense, the system enters a superconductor state with one
Goldstone mode. This superconductor is the triplet superconductor
TSC$_z$ in the matrix order parameter Eq.~\ref{Q}. If we define an
O(4) vector $\vec{\phi} = (\mathrm{Neel}_x, \mathrm{Neel}_y,
\mathrm{Re}[\mathrm{TSC}_z], \mathrm{Im}[\mathrm{TSC}_z])$, the
nonlinear sigma model of $\vec{\phi}$ has a topological
$\Theta-$term. A N\'{e}el-TSC transition on the honeycomb lattice
was discussed in Ref.~\cite{ran2008}. However, this N\'{e}el-TSC
transition does {\it not} happen in the KMH model, since in the
KMH model $z^c_1$ and $z^c_2$ are nondegenerate.

\subsection{Liquid-N\'{e}el transition}

In the KMH model, the phase transition between the spin-charge
liquid phase and the N\'{e}el phase does not involve any low
energy charge degrees of freedom, thus this transition can be
understood as the condensation of $z^s_\alpha$, while $z^c_\alpha$
are gapped. When $z^c_\alpha$ are gapped out, they can be safely
integrated out from Eq.~\ref{lcs1}, then $z^s_1$ and $z^s_2$
become degenerate. $z^s_\alpha$ is coupled to a $Z_2 \times Z_2 $
gauge field, as was discussed in Ref.~\cite{xu2010,xuludwig}. Now
the liquid-N\'{e}el transition is described by the following
Lagrangian \beqn \mathcal{L} &=& \sum_{\alpha = 1}^2 |\partial_\mu
z^s_\alpha|^2 + r_s|z^s_\alpha|^2 + g(\sum_{\alpha = 1}^2
|z^s_\alpha|^2)^2 \cr\cr &+& u |z^s_1|^2|z^s_2|^2 + \cdots
\label{liquidneel}\eeqn The N\'{e}el order parameter is a bilinear
of $z^s_\alpha$: \beqn N^x \sim \mathrm{Re}[(z^s)^t i\sigma^y
\sigma^x z^s] \sim \mathrm{Re}[z_1^2 - z_2^2], \cr\cr N^y \sim
\mathrm{Re}[(z^s)^t i\sigma^y \sigma^y z^s] \sim
\mathrm{Re}[iz_1^2 + iz_2^2]. \eeqn


The first line of Eq.~\ref{liquidneel} has a full O(4) symmetry,
while the second line breaks the O(4) symmetry down to
$\mathrm{U(1) \times U(1)} \times Z_2$. In this field theory $u >
0$, thus in the condensate of $z^s_\alpha$ there is only one
Goldstone mode that corresponds to the inplane N\'{e}el order.
According to the high order $\epsilon$ expansion in
Ref.~\cite{vicari2003b}, $u$ is a relevant perturbation at the 3d
O(4) universality class, which is expected to drive the transition
first order eventually.

\subsection{The $s-$wave superconductor}

The phase $A4$ in Fig.~\ref{honeypd} is reduced to the $s-$wave
superconductor in Fig.~\ref{KMHphase}$a$. The $s-$wave SC is the
charge-dual of the inplane N\'{e}el order in phase $A2$. Just like
the magnetic vortex of the N\'{e}el order, the vortex of the SC
also carries two quantum numbers: spin-$\pm1/2$ and vorticity:
$(s, v)$. The symmetry of the KMH model divides the vortices into
two groups: $(\frac{1}{2}, v)$ and $(-\frac{1}{2}, -v)$ are
degenerate, while $(\frac{1}{2}, - v)$ and $(- \frac{1}{2}, v)$
are degenerate. These two groups of vortices are precisely
$(z^s_1, z^{s \ast}_1)$ and $(z^s_2, z^{s \ast}_2)$ introduced in
Eq.~\ref{lcs}.

The quantum phase transition between the $s-$wave SC (phase $A4$)
and the QSH insulator (phase $A$) is interpreted as condensing
either $z^s_1$ or $z^s_2$, this transition is a 3d XY transition.
The transition between the liquid phase $A3$ and phase $A4$ is
described by a similar theory as Eq.~\ref{liquidneel}, and it is
expected to be a first order transition.

\subsection{Multicritical point}

There is a multicritical point in our phase diagrams
Fig.~\ref{honeypd} for SO(4) invariant systems, which separates
the liquid phase from the QSH insulator. In
Ref.~\cite{xusachdevtriangle}, this multicritical point was
studied using a large$-N$ generalization, where $N$ is the number
of components of $z^s_\alpha$ and $z^c_\alpha$. It was
demonstrated that when $N$ is large enough, this multicritical
point is a conformal field theory \cite{xusachdevtriangle}.
Compared with the theory studied in Ref.~\cite{xusachdevtriangle},
Eg.~\ref{lcs1} has several SO(4) symmetry breaking perturbations,
for example the last term in Eq.~\ref{lcs1}. If we extrapolate the
$1/N$ calculation in Ref.~\cite{xusachdevtriangle} to our current
case with $N = 2$, the last term in Eq.~\ref{lcs1} is a relevant
perturbation at the multicritical conformal field theory. Thus we
expect the transition between the liquid and the QSH insulator in
the KMH model to be a first order transition.

\section{Compare with other theories}

Based on our theory discussed in this paper, we propose the phase
diagram for the KMH model in Fig.~\ref{KMHphase}$b$ plotted
against $ \lambda$ and $U$. When $\lambda = 0$, there is one extra
transition inside the N\'{e}el phase, which corresponds to the
transition between the pure N\'{e}el order with GSM $S^2$, and the
N\'{e}el + nematic order with GSM $\mathrm{SO(3)/Z_2}$ (phase $A2$
in Fig.~\ref{honeypd}). This transition is {\it absent} once
nonzero $\lambda$ is turned on, this is because the symmetry of
the N\'{e}el + nematic order order is identical to the N\'{e}el
order with a background QSH spin-orbit coupling.

In Ref.~\cite{ran2010b}, the authors proposed a different phase
diagram for the Hubbard model on the honeycomb lattice. Instead of
a N\'{e}el+nematic order, the authors of Ref.~\cite{ran2010b}
predicted a chiral antiferromagnetic (CAF) phase with an extra
nonzero order $\langle \nu_{ij} \hat{z}\cdot(\vec{S}_i \times
\vec{S}_j) \rangle \neq 0$ between next nearest neighbor sites.
Unlike the N\'{e}el+nematic order, the CAF order has a different
symmetry from the pure N\'{e}el order even with presence of the
QSH spin-orbit coupling in the background. For instance, when the
N\'{e}el vector is along the $y$ direction, it is still invariant
under the PH transformation defined in Eq.~\ref{PH}; but the CAF
order breaks this PH transition, as was discussed in section IIIB.

This symmetry analysis leads to the following two conclusions:

{\it (1).}  If there is a CAF phase with zero $\lambda$, there
must be a transition line within the inplane N\'{e}el phase in
Fig.~\ref{KMHphase}$b$ even with finite $\lambda$ (dashed line of
Fig.~\ref{KMHphase}$b$), which corresponds to the order-disorder
transition of the CAF order parameter.

{\it (2).} The transition between the QSH insulator and the
N\'{e}el+CAF phase is {\it not} a 3d XY transition, because the
symmetry breaking at this transition is different from the 3d XY
transition.

The different predictions between our theory and
Ref.~\cite{ran2010b} can be checked numerically in the future.

\begin{acknowledgements}

The authors thank YoungHyun Kim for very helpful discussions.
Cenke Xu is supported by the Alfred P. Sloan Foundation. Christian
Griset is supported by the NSF GRFP Fellowship DGE-0707460.

\end{acknowledgements}

\bibliography{QSHH}

\end{document}